\magnification=\magstep1
\nopagenumbers
\centerline{A Java Calculator of Standard Big Bang Nucleosynthesis}
\medskip
\centerline{Luis Mendoza and Craig J. Hogan, University of Washington}
\bigskip
\centerline{Abstract}
A simple Java applet is presented which allows quick and easy computation and plotting of
the predictions of light element abundances in SBBN,
including errors in the predictions propagated from input reaction rates
correlated and  calibrated via Monte Carlo.
The applet, which requires Java 1.1-compatible browsers,
 can be found at

\noindent
 http://www.astro.washington.edu/research/bbn/ 
\bigskip

Standard Big Bang Nucleosynthesis refers to an exceptionally simple and beautiful
 model of the early universe. Matter and radiation are
spatially uniform and the expansion is controlled by known fields
 with negligible chemical potentials, so the dynamics of the expansion and
density as a function of time and temperature is determined by ``just physics''
 with no external parameters. The production of light elements
depends on one parameter, the small baryon excess
 characterised by the baryon-to-photon ratio $\eta$. Numerical integration of the reaction
network yields predictions for light element abundances with ``theoretical 
errors'' dominated by the experimental errors in the input physics
such as nuclear reaction rates. The error matrix for SBBN has been computed
 by Fiorentini et al. (Phys. Rev. D58,1998; astro-ph/9803177)
by Monte Carlo techniques. 

For many years SBBN theory and observation were mainly compared by
 the theorists. Integrations using updated rates and techniques
were used to make state-of-the-art predictions and these were then
 compared with a collection of disparate observational data. Many
reviews of this kind successively refined the predictions and the
 arguments concerning the post-big-bang processing of the elements (see
reviews by S. Sarkar, Rep. Prog. Phys. 59, 1493, 1996, and D. Schramm
 and M. S. Turner, Rev. Mod. Phys. (Colloquia) 70, 303, 1998).
The 
theoretical errors
are no longer negligible compared with the observational errors so
correct statistical comparison requires handling all of the errors at the same time, 
forcing the theorists to interpret observers' error budgets
or vice versa. With theoretical predictions and errors now stable, it makes sense to 
empower the observers to do the comparison. 

This applet provides easy access to the precise predictions of SBBN including calibrated errors,
 as described by Fiorentini et al. (1998). This
version (1.1) includes the predictions and errors for deuterium, helium-3, helium-4, and 
lithium-7 abundances. It will be useful in using data
on each abundance (with errors) to set statistical limits on $\eta$, or to make predictions
 for abundances from other estimates of $\eta$ (e.g., M.
Fukugita, C. J. Hogan and P. J. E. Peebles, ApJ 503, 518, 1998). 
Future versions will allow statistical concordance tests of SBBN and estimates
 of $\eta$ using more than one abundance datum, as well as
allowing changes in the input reaction rates and errors.

\bye